\documentclass[prl,showpacs,twocolumn]{revtex4}

\usepackage{graphicx}
\usepackage{dcolumn}
\usepackage{bm}

\begin{document}
\catcode`\ä = \active \catcode`\ö = \active \catcode`\ü = \active
\catcode`\Ä = \active \catcode`\Ö = \active \catcode`\Ü = \active
\catcode`\ß = \active \catcode`\é = \active \catcode`\è = \active
\catcode`\ë = \active \catcode`\ô = \active \catcode`\ê = \active
\catcode`\ø = \active \catcode`\ò = \active \catcode`\í = \active
\defä{\"a} \defö{\"o} \defü{\"u} \defÄ{\"A} \defÖ{\"O} \defÜ{\"U} \defß{\ss} \defé{\'{e}}
\defè{\`{e}} \defë{\"{e}} \defô{\^{o}} \defê{\^{e}} \defø{\o} \defò{\`{o}} \defí{\'{i}}

\preprint{APS/123-QED}

\title{Metastable Bose-Einstein Condensate in a Linear Potential}

\author{D. S. Naik, S. R. Muniz and C. Raman}
 \email{craman@gatech.edu}
\affiliation{%
School of Physics, Georgia Institute of Technology, Atlanta,
Georgia 30332
}%
\date{\today}
\begin{abstract}
We have created a Bose-Einstein condensate whose spin orientation is
metastable.  Condensates were transferred into a quadrupole magnetic
trap, where Majorana transitions limited the lifetime to a few
hundred milliseconds, about 30 times the trapping period.  Atoms
held in the trap frequently displayed a ring-shaped time-of-flight
distribution. We speculate that such a ring could be either a
quantized vortex or a feature of the Majorana loss dynamics in the
quantum regime.
\end{abstract}

\pacs{03.75.Mn, 03.75.Lm, 03.75.Nt, 32.80.Lg, 32.80.Pj}
\maketitle                                

Matter has many states which are fascinating, but whose existence
is only transient.  Classic examples of metastability in
macroscopic quantum systems are the ``persistent'' currents of
superfluids and superconductors \cite{till90}. Quantum gases have
been shown to exhibit a rich variety of metastable states,
including spatial spin domains in Bose-Einstein condensates (BECs)
\cite{mies99meta}, lattices of quantized vortices
\cite{abos01latt}, as well as the molecular states populated in
the vicinity of a Feshbach resonance \cite{donl02}.
In all these examples, the lifetime depends
critically on atom-atom interactions.  In this work, we report on
the experimental observation of a trapped BEC whose spin
orientation is metastable in an external, inhomogeneous magnetic
field.  The decay of the trapped condensate thus depends primarily
on {\em single} atom physics.  The trapping field can in principle
be used to coherently control the coupling between spin and
spatial wavefunctions.

We produced the metastable BEC by transferring atoms into a magnetic
quadrupole trap (see Figure 1).  This is the first realization of a
three-dimensional linear potential for BECs.  Due to the presence of
the magnetic field zero at the center, atoms spontaneously flipped
their spin and entered an untrapped state.  However, this process was
relatively slow, and our condensate could remain trapped for
hundreds of milliseconds, about 30 times the characteristic
oscillation time in the linear trap.  Moreover, we frequently
observed that atoms released from the trap had a ring-shaped
time-of-flight distribution reminiscent of the expansion of vortex
states.  We speculate on possible mechanisms for the formation of
this ring.

An atom moving through an inhomogeneous $B$-field can flip its spin through a nonadiabatic process. Such Majorana transitions are most significant when the magnitude of $\vec{B}$ is close to zero, as it is near a region of radius $b \sim 1 \mu$m dimension near the center of a quadrupole trap.  If the de Broglie wavelength $\lambda \leq b$, as it is usually for thermal atoms, one may use a semiclassical picture where the atomic motion can be treated classically and the internal spin states are treated quantum-mechanically.  For a BEC, however, one must treat the entire problem quantum mechanically since all atoms occupy a single wavefunction that extends over a region that is typically much larger than $b$.  For atoms whose internal spin is $F$ and a three-dimensional quadrupole magnetic field $\vec{B} = B'(x \hat{x} + y \hat{y} - 2 z \hat{z})$, the resulting Hamiltonian contains the potential energy term:
\begin{equation}
V = -\vec{\mu}\cdot \vec{B}(\vec{r}) = \mu_B g_F B'(x \hat{F}_x + y
\hat{F}_y - 2 z \hat{F}_z)
\end{equation}
where $\vec{F} = (\hat{F}_x,\hat{F}_y,\hat{F}_z)$ is the vector spin-$F$ operator and $\mu_B$ is the Bohr magneton. Since $V$ acts on both spin and spatial coordinates, the trapped and untrapped states are coupled to one another.  In general, this process is completely coherent, and could be used as a mechanism for creating spinor condensates, as has been discussed for other inhomogeneous field geometries \cite{pu01}.  We note that the same coupling exists in Ioffe-Pritchard traps, but is much smaller due to the finite bias field $B_z$ that preserves the spin orientation.
\begin{figure}[htbp]
\begin{center}
\vspace{-2mm}
\includegraphics[width = 0.5\textwidth]{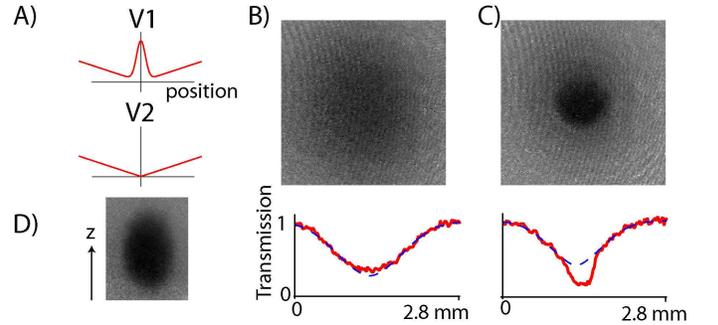}
\end{center}
\vspace{-7mm}\caption{(Color online). Metastable Bose-Einstein condensate.  (A)
Atoms are transferred from an optically plugged trap
(potential V1) into a purely linear trap (potential V2) by slowly
reducing the plug intensity to zero.  (B) Transfer of purely
thermal atoms cooled to a final rf value of 0.35 MHz, and (C)
transfer of Bose-Einstein condensates at 0.25 MHz rf.  Each image
was taken along the $\hat{z}$ axis after 20 ms TOF, and is 1.6
$\times$ 1.6 mm. Below each image is a slice through the data as
well as a fit to the thermal wings (dashed lines). (D) View
perpendicular to the $\hat{z}$ axis shows anisotropic expansion of
the BEC after 15 ms TOF. Image is 0.5 $\times$ 0.7 mm.}
\label{fig:fig1} \vspace{-2 mm}
\end{figure}

We begin by creating Bose-Einstein condensates of about $5 \times
10^6$ sodium atoms in the $F=1, M_F=-1$ state in an optically
plugged quadrupole trap (OPT) as described in \cite{naik05}.
Majorana losses are suppressed by focusing a 40 $\mu$m, far-off
resonance, blue-detuned, 532 nm ``plug'' laser beam onto the
spatial location of the magnetic field zero.  After making a BEC,
we ramp the intensity of the plug laser from an AC Stark shift of
60 $\mu$K to zero over 200 ms, leaving the condensate trapped by a
pure quadrupole magnetic field whose axial field gradient could be
varied.  After a variable hold time, we shut off the quadrupole
coils allowing for times-of-flight up to 35 milliseconds.  Five to
ten milliseconds before imaging, we turned on a $\hat{z}$-directed
bias field of approximately 1 Gauss. Absorption imaging was
performed on either the $F=1$ to $F'=2$ transition using a 150
$\mu$s probe pulse or by first optically pumping the atoms into
the $F=2$ ground state with a 1.5 to 4 ms pump pulse followed by a
150 $\mu$s probe pulse on the $F=2$ to $F'=3$ cycling transition.

Figure 1 encapsulates our main result:  we have produced a
condensate in the purely linear potential of the quadrupole coils.
That it merely exists is rather counter-intuitive, since one might
expect the atoms to rapidly leave the trap by Majorana
transitions.  In fact, as we show below, these transitions occur
slowly when compared with the natural timescales of the system. We
observed bimodal time-of-flight distributions (Figure 1c) when
atoms were imaged along the symmetry axis of the quadrupole field
$\hat{z}$ (which coincides with the direction of gravity).  Due to
the presence of the condensate, the fit to the thermal wings does
not agree with the data in the center of the cloud.  The fit
agreed well, however, when we transferred a purely thermal cloud,
as in Figure 1b.  The BEC expanded anisotropically when viewed
from the side (Figure 1d) due to the increased mean-field pressure
arising from the factor of 2 larger axial magnetic field gradient
compared with the radial ($\hat{x}$ or $\hat{y}$) direction.

Unlike a condensate in the Thomas-Fermi regime in a harmonic
potential, the time-of-flight expansion does not obey a simple
scaling of the parabolic distribution.  In some cases, the entire
shape of the distribution had changed dramatically, as we will
discuss later.  Therefore, we used a model-independent approach to
analyze the images and extract both condensate number and thermal
number.  We removed the central region that contained mostly the
condensate and fit a 2-dimensional thermal distribution to the
remaining image. We deduced the condensate number by subtracting
the total thermal number obtained from the fit from the total atom
number obtained by summing the column density over the entire
original image. The total number was adjusted for shot-to-shot
background level fluctuations using the background obtained from
the thermal fit.

\begin{figure}[htbp]
\begin{center}
\includegraphics[width = 0.47 \textwidth]{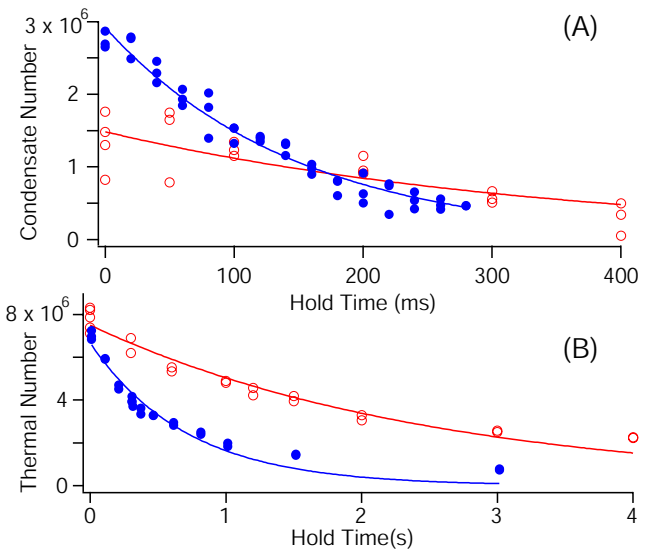}
\end{center}
\vspace{-7mm}\caption{(Color online). Decay in the linear trap.  (A) Condensate
number is plotted against hold time for traps with axial magnetic
field gradients of 21 G/cm (filled circles) and 13 G/cm (open
circles). (B) Decay of purely thermal clouds with field gradients
of 21 G/cm (open circles) and 105 G/cm (filled circles),
corresponding to measured temperatures of $2.7 \mu$K and $8.2
\mu$K, respectively. The solid lines are fits to a single
exponential, as described in the text.} \label{fig:fig2}
\vspace{-5mm}\end{figure}

We observed our condensate to have a lifetime of a few hundred milliseconds in the purely linear potential.  We attribute the decay to Majorana transitions to an untrapped state, which should occur in a narrow region near the cloud center. By comparison, in an rf-shielded OPT, without Majorana loss, condensates were observed for more than 12 seconds. Figure 2a shows the decay curve for traps with two different axial magnetic field gradients, 21 G/cm and 13 G/cm, along with a fit to a single exponential decay for each. The condensate fractions were 45 \% and 30 \%, respectively, at zero hold time.  The measured decay rates were 7 s$^{-1}$ and 3 s$^{-1}$, respectively. We can compare this rate to a WKB estimate of the oscillation frequency which in general depends on the size of the condensate \cite{merz_book}.  The result is $\nu_{osc} = \frac{1}{4 \sqrt{2}} \sqrt{\frac{\tilde{\mu}B'}{M R}}$, where $\tilde{\mu} = \frac{1}{2} \mu_B$ is the magnetic moment and $B' = \frac{2}{3} B_z'$ is the spatially averaged field gradient. $R$ is the average condensate radius, which we estimate in the section below to be $\frac{3}{4} R_\rho = 13.5 \mu$m for our 21 G/cm trap, where $R_\rho$ is the Thomas-Fermi radius in the $x-y$ plane. This yields $\nu_{osc} = 200$ Hz, about 30 times faster than the rate of decay. Such a long lifetime was only possible by using such weak field gradients.  In the case of the weaker trap with 13 G/cm gradient, the loss rate was smaller due to the larger condensate size, as discussed below. Indeed, we could observe a BEC for up to 0.5 s in this trap. Longer lifetimes could not be obtained by weakening the trap much further due to the difficulty of balancing gravity, which requires at least 8 G/cm.

One may use a semiclassical model to estimate the Majorana rate,
as in reference \cite{petr95}.  In this model, the rate of loss
$\Gamma_M$ only depends on the cloud size,
\begin{equation}
\Gamma_M \sim \frac{\hbar}{M R^2} \label{eq:two}
\end{equation}
and is simply related to the statistical probability that an atom
trapped in a cloud of radius $R$ will pass directly through a hole
of radius $b = \sqrt{\frac{\hbar v}{\tilde{\mu} B'}}$ located at the
origin.  For a gas above the transition temperature, the scaling
relation (\ref{eq:two}) was verified in previous work \cite{petr95}.
To compare with our system, we measured the loss rate for thermal
clouds just above the transition temperature. The data are shown in
Figure 2b for field gradients of 21 G/cm and 105 G/cm, corresponding
to temperatures of $2.7 \mu$K and $8.2 \mu$K and average cloud sizes
of $R = \frac{k_B T}{\tilde{\mu}{B'}} = 56 \mu$m and $35 \mu$m,
respectively.  We measured the loss rate by fitting the data to an
exponential decay, which should be accurate at least for short times
before the loss causes the cloud to heat up significantly.  The
theoretical prediction from Eqn. (\ref{eq:two}) for a gas at the
same temperatures are $0.9$ s$^{-1}$ and $2.2$ s$^{-1}$, while we
measured $0.4$ and $1.4$ s$^{-1}$, respectively. This gives us
confidence that the semiclassical scaling of Eqn. (\ref{eq:two})
yields the observed loss rate within a factor of about 2.

For a Bose-Einstein condensate, the spin-flip occurs through a {\em
coherent} process governed by the Gross-Pitaevskii (GP) equation for
the wavefunction $\Psi$, which is a $2F+1$ component spinor defined
with respect to the space-fixed $\hat{z}$-axis. Solving this
equation is in general quite difficult due to the coupling between
spin and space which was discussed earlier.  However, we gain
considerable insight by viewing the problem in a spatially varying
spin basis defined with respect to the {\em local} magnetic field
direction \cite{berg89,metc99}. We apply a local transformation
$\Lambda = \Lambda(\vec{r})$ that diagonalizes the potential energy
term $V$ in spin space. The resulting GP equation is
\begin{equation}
\left[\Lambda \left(\frac{\hat{p}^2}{2 M}\right) \Lambda^{-1}
\right]\Phi + \left[ g n + V_d -\mu \right]\Phi =
0\label{eq:three}
\end{equation}
where $\mu$ is the chemical potential, $\Phi = \Lambda \Psi$, and
$V_d = \Lambda V \Lambda^{-1}$ is diagonal in the local spin basis.
Its eigenvectors are the weak-field seeking , strong-field seeking
and field-insensitive states, with eigenvalues
$+\tilde{\mu}|\vec{B}(\vec{r})|$, $-\tilde{\mu}|\vec{B}(\vec{r})|$
and $0$, respectively.  The trapped state is the weak-field seeking
state.  The nonlinear interaction term is $g n$, where $n$ is the
particle density, $g = 4 \pi \hbar^2 a/M$ and $a$ is the two-body
scattering length.  We have neglected the spin dependence of this
term.  This is a valid assumption in our case since we are only
interested in the evolution of the trapped state, upon which the low
density untrapped particles should have little effect.

The second term of Eqn. (\ref{eq:three}), $[gn+V_d-\mu]$, is
diagonal in the local spin basis and therefore describes a set of {\em
adiabatic} potentials determined by the eigenvalues of $V_d$.
However, since the kinetic energy operator
does not commute with $\Lambda(\vec{r})$, the first term couples
the different spin channels together, resulting in Majorana transitions.

One must solve the problem using a coupled channel approach similar
to that involved in the theory of Feshbach resonances \cite{fried90}.
While such a solution is beyond the scope of this paper,
we can compare the loss rate for our BEC using the semi-classical
Eqn. (\ref{eq:two}).  We use the Thomas-Fermi density distribution
for the linear potential, which is a cone-shaped function: $n(x,y,z)
= n_0 (1-\frac{1}{R_\rho}\sqrt{x^2+y^2+4 z^2})$, where $n_0$ is the
peak density.  We can determine $n_0$ self-consistently using the
relations $g n_0 = \mu$ and $\mu = \tilde{\mu} B'_z R_\rho/2$ where
$\mu$ is the chemical potential. For our 21 G/cm trap, $\mu \simeq
0.65 \mu$K yielding $R_\rho = 18 \mu$m for $N = 3 \times 10^{6}$
atoms.  Taking the average radius to be $R=\frac{3}{4} R_\rho$ as we
have earlier for the thermal cloud, we estimate that the condensate
should decay at a rate which is $(56/13.5)^2 = 17$ times faster than
the thermal cloud in the same trap.  Using the experimental value of
$0.4$ s$^{-1}$ from Figure 2b, we get $\Gamma_M = 6.9$ s$^{-1}$,
which compares fortuitously well with our measurement of 7 s$^{-1}$.
We conclude that the loss dynamics for both condensed and normal
components are fairly similar, when the cloud sizes are scaled
appropriately.

\begin{figure}[htbp]
\begin{center}
\vspace{-4mm}
\includegraphics[width = 0.48\textwidth]{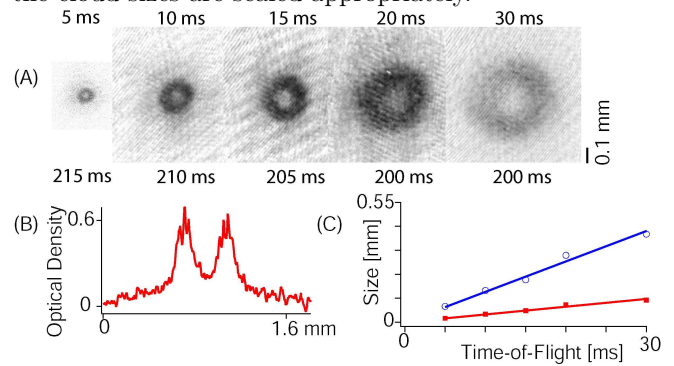}
\end{center}
\vspace{-6mm}\caption{(Color online). Ring shaped expansion of a BEC.  (A) Images
of a condensate after various hold times (given below each image)
and times of flight (given above). (B) Slice through the data at
30 ms TOF shows a high-contrast density minimum on top of the
thermal background. (C) Radius of the central hole (filled
squares) and the condensate (open circles) versus time-of-flight .
The velocities of expansion are 14 mm/sec and 3.5 mm/sec,
respectively.} \label{fig:fig3} \vspace{-3mm}\end{figure}

While the loss rate is fairly well understood, the quantum dynamics of the trapped atoms are more complex.  We have observed new, intriguing features of the condensate in the linear trap for which there is no current explanation, although a more detailed investigation is ongoing. After hold times of 100-200 ms, the time-of-flight distribution frequently displayed a dramatic and unexpected signature--a single, clear hole in the center of the cloud.  This hole reflects a depletion of low momentum atoms in the image. Figure 3a shows a succession of images taken for different times-of-flight. We used $F=2$ imaging, which is spin-state independent.  This established that the observed ring feature is a variation in the {\em total} atomic density and not due to the formation of possible spin domains that might have different optical absorption \cite{extra2}.  Accounting for the thermal background in the image, the density minimum in the center was typically $50$\% of the condensate peak density, but on some occasions could be about $90$\%.  The latter case is shown in Figure 3b. Ring-shaped condensates were visible in approximately half of the experimental runs, while for the remaining images the absorption profile was relatively smooth and showed little to no structure. Moreover, we could not easily see the ring in the images with no hold time.  It was possibly obscured by high optical density. Similar results were obtained for traps of both 21 G/cm and 13 G/cm, indicating it is a general feature of the linear potential.

We observed that the ring structure had appeared after only 5 ms
of expansion. For each image in Figure 3a, we varied the detuning
of the optical pumping light to control the number of atoms pumped
into the $F=2$ state for imaging.  This resulted in an optical
depth close to 1 for each image without introducing spatial
distortions. In Figure 3c, we have plotted the hole radius as well
as the overall cloud radius as a function of the time-of-flight.
To quantify these parameters, we fit the ring pattern from a slice
such as shown in Figure 3b to the sum of 2 Gaussians at locations
$x_1,x_2$ with $1/e$ half-widths $w_1,w_2$.  The condensate radius was
defined to be $R_c = \frac{1}{2} (x_2-x_1+w_1+w_2)$, while the hole
radius was $R_h = \frac{1}{4}(x_2-x_1)$.  Both $R_h$ and
$R_c$ are plotted in Figure 3c and increase linearly with time,
with $\frac{d R_h}{d t}$a factor of 4 smaller.

Our observation of the ring pattern is unexpected, and we do not
have a conclusive explanation for it as of yet.  However, our data
in Figure \ref{fig:fig3} clearly shows that the wavefunction of the
condensed state before expansion has a complex structure.  The ring
feature we observe could be a vortex state--the core of a vortex
should expand at a velocity which is roughly proportional to that of
the overall condensate size \cite{lund98}, consistent with our
observations.  There could be two separate condensates formed in the optically plugged trap which are merged together when the plug is removed \cite{naik05}.  Phase fluctuations between the
two uncorrelated BECs could create topological defects, including vortices.
The random character of this process might explain why we do not observe
the hole on every cycle of the experiment.

The stability of a vortex in a linear potential has not been
explored theoretically.  A rough estimate suggests, however, that
the presence of a vortex should not interfere with the spin-flip process
occurring near the trap center. This is due to the fact that the depletion of the
superfluid density occurs only within a region of the size of the vortex core,
typically $\approx 0.13 \mu$m for our system, considerably smaller than the size of
the hole, $b = \sqrt{\frac{\hbar c}{\tilde{\mu} B'}} \approx 1.5 \mu$m, estimated from the semiclassical model \cite{petr95} replacing the atomic velocity with that of sound propagation for our condensate, $c =$ 1.5 cm/s.  Indeed, we observed no correlation between observation of the hole and increased lifetime of the BEC.

One unique feature of the linear potential is that the magnetic
field direction varies in space, unlike most Ioffe-Pritchard traps.
Therefore, the condensate has an {\em inhomogeneous} spin
orientation, and it is plausible that a vortex state could be
created by the shut-off of the trap, when the inhomogeneous field is
removed.  We are investigating this mechanism further.

The {\em spatially localized} character of Majorana loss may also
play a role in creating the pattern observed in Figure
\ref{fig:fig3}. Indeed, the time evolution is quite rich, as the
initially stationary condensate begins to acquire an inward velocity
field toward the magnetic field zero to balance the loss. Spatial
patterns have been observed in electromagnetic wave propagation
\cite{porr04} and trapped BECs with negative scattering length
\cite{ueda03}, both examples of systems where nonlinearity
effectively confines losses to a small region of space where the
wave amplitude is highest.  They bear a strong resemblance to our
trapped BEC, where although the losses are {\em linear} in atom
density, they are localized due to the magnetic field. Numerical
simulations are necessary to shed more light on the dynamics of this
novel state.

In summary, we have demonstrated metastable Bose-Einstein
condensates in a purely linear potential.  Majorana losses are
shown to have intriguing effects on the time-of-flight evolution.
These could be signatures of vortex formation, which could be
verified, for example, by interferometric means \cite{inou01vort}.

We thank Li You, T. A. B. Kennedy, P. di Trapani, M. S. Chapman
and M. Bhattacharya for useful discussions.  This work was
supported by the DoE, ARO and by Georgia Tech.


\begin{thebibliography}{16}
\bibitem[{ext()}]{extra2}
\bibinfo{note}{In any case, we do not see any evidence of untrapped atoms in
  the images due to their low density.}
\end{thebibliography}

\end{document}